\documentclass{pasj00}
\twocolumn
\draft
\begin{document}

\SetRunningHead{Naik et al.}{Suzaku Observation of CXOU~J164710.2--455216}

\title{Suzaku Observation of the Anomalous X-ray Pulsar CXOU~J164710.2--455216}
\author{Sachindra \textsc{Naik}\thanks{Present address: Physical
Research Laboratory, Navrangpura, Ahmedabad - 380 009, India}\altaffilmark{1},
Tadayasu \textsc{Dotani}\altaffilmark{1,2,3},
Nobuyuki \textsc{Kawai}\altaffilmark{3},
Motohide \textsc{Kokubun}\altaffilmark{1},\\
Takayasu \textsc{Anada}\altaffilmark{1},
Mikio \textsc{Morii}\altaffilmark{4},
Tatehiro \textsc{Mihara}\altaffilmark{5},
Teruaki \textsc{Enoto}\altaffilmark{6},\\
Madoka \textsc{Kawaharada}\altaffilmark{6}
Toshio \textsc{Murakami}\altaffilmark{7},
Yujin E. \textsc{Nakagawa}\altaffilmark{8},\\
Hiromitsu \textsc{Takahashi}\altaffilmark{9},
Yukikatsu \textsc{Terada}\altaffilmark{5},
Atsumasa \textsc{Yoshida}\altaffilmark{8}
}
\altaffiltext{1}{Institute of Space and Astronautical Science, JAXA, 
     3-1-1 Yoshinodai, Sagamihara,\\ Kanagawa 229-8510, Japan}
\email{naik@astro.isas.jaxa.jp}
\altaffiltext{2}{Space and Astronautical Science, School of Physical Sciences,\\
     The Graduate University for Advanced Studies,\\
     3-1-1 Yoshinodai, Sagamihara, Kanagawa 229-8510, Japan}
\altaffiltext{3}{Department of Physics, Tokyo Institute of Technology,\\
2-12-1 Okayama, Meguro-ku, Tokyo 152-8551, Japan}
\altaffiltext{4}{Department of Physics, Rikkyo University, 3-34-1, 
    Nishi-Ikebukuro, Toshima-ku,\\ Tokyo 171-8501, Japan}
\altaffiltext{5}{Cosmic Radiation Laboratory, Institute of Physical and 
    Chemical Research, Wako, \\ Saitama 351-0198, Japan}
\altaffiltext{6}{Department of Physics, University of Tokyo, 
7-3-1 Hongo, Bunkyo-ku, \\ Tokyo 113-0033, Japan}
\altaffiltext{7}{Department of Physics, Kanazawa University, Kadoma, \\
Kanazawa, Ishikawa 920-1192, Japan}
\altaffiltext{8}{Graduate School of Science and Engineering, Aoyama
Gakuin University, Sagamihara, \\ Kanagawa 229-8558, Japan}
\altaffiltext{9}{Department of Physical Science, Hiroshima University,
1-3-1 Kagamiyama, \\ Higashi-Hiroshima, Hiroshima 739-8526, Japan}

\maketitle

\begin{abstract}
Suzaku TOO observation of the anomalous X-ray pulsar CXOU~J164710.2-455216 
was performed on 2006 September 23--24 for a net exposure of 38.8 ks. During the 
observation, the XIS was operated in 1/8 window option to achieve a time 
resolution of 1 second. Pulsations are clearly detected in the XIS light 
curves with a barycenter corrected pulse period of 10.61063(2)~s. The XIS 
pulse profile is found to be highly non-sinusoidal. It shows 3 peaks of 
different amplitudes with RMS fractional amplitude of $\sim$11\% in
 0.2--6.0~keV energy band. 
Though the source was observed with the Hard X-ray Detectors (HXD) of 
Suzaku, the data is highly contaminated by the nearby bright X-ray source 
GX~340+0 which was in the HXD field of view. The 1--10~keV XIS spectra 
are well fitted by two different models consisting of a power-law and a
blackbody component and two blackbody components respectively. Although both
the models are statistically acceptable, difference in the pulse profiles at
soft (0.2--6.0~keV) and hard (6--12~keV) X-rays favors the model consisting
of two blackbody components. The temperatures of two blackbody
components are found to be 0.61$\pm$0.01~keV and 1.22$\pm$0.06~keV and the 
value of the absorption column density is 1.73$\pm$0.03 $\times$ 10$^{22}$ 
atoms cm$^{-2}$. The observed source flux in 1--10~keV energy range is
 calculated to be 2.6 $\times$ 10$^{-11}$ ergs cm$^{-2}$ s$^{-1}$ with
 significant contribution from the soft blackbody component 
 ($kT$ = 0.61~keV). 
Pulse phase resolved spectroscopy of XIS data shows that the flux of the soft 
blackbody component consists of three narrow peaks, whereas the flux of the 
other component shows a single peak over the pulse period of the AXP\@. 
The blackbody radii changes between 2.2--2.7~km and 0.28--0.38~km 
(assuming the source distance to be 5~kpc) over pulse phases for the
 soft and hard components, respectively. 
The details of the results obtained from the timing and 
spectral analysis is presented.
\end{abstract}

\section{Introduction}

The Anomalous X-ray pulsars (AXPs) are a small group of X-ray pulsars 
which show common properties such as (i) known to present 
within 1$^\circ$ of the Galactic plane, (ii) pulse period in a very narrow 
range of 5--12~s, unlike the radio pulsars and accreting X-ray pulsars, 
(iii) large and more or less steady spin-down or braking, in contrast to most
accretion powered pulsars which show spin ups and downs, (iv) similar spectra 
with soft component characterized by a blackbody model with temperature below
1 keV, with additional harder component in some cases, and (v) relatively high 
X-ray luminosity ($\sim$10$^{34}$--10$^{36}$~erg~s$^{-1}$) which cannot 
be obtained from the loss of rotational energy of a neutron star alone.
Some of the AXPs also emit at optical and/or infrared wavelengths 
(Hulleman et al.\ 2000; Durant \& van Kerkwijk 2006b).
The AXPs are considered to be very young (10$^{3}$--10$^{5}$~yr), some
of which are associated with supernova remnants (SNRs). 
The AXPs are considered to 
be the neutron stars with the strongest known magnetic field 
(10$^{14}$--10$^{15}$~G), though a direct measurement to confirm 
the presence of such high magnetic field strengths is still needed. 
The source of energy for the radiative emission in AXPs is described 
by the magnetar model, in which the decay of an ultra-strong magnetic 
field powers the high-luminosity bursts and also a substantial fraction 
of the persistence X-ray emission (Thompson \& Duncan 1996). The 
competing model for the mechanism of powering the X-ray emission 
in AXPs is that the AXPs are neutron stars surrounded by fossil disks that 
were acquired during supernova collapse or during a common-envelope 
interaction (Corbet et al.\ 1995; van Paradijs et al.\ 1995; Chatterjee \& 
Hernqyist 2000). The properties of AXPs are summarized by Mereghetti et al.\
(2002), Kaspi \& Gavriil (2004), Woods \& Thompson (2004). 

Recently discovered AXP CXOU~J164710.2-455216 is located in the young, massive 
Galactic star cluster Westerlund 1 (Muno et al.\ 2006). Coherent pulsations
with a period of 10.61 s were detected from the 2005 May 22 and June 18 
$Chandra$ observations of the cluster. The best-fit periods obtained from 
above two observations were 10.6112(4)~s and 10.6107(1)~s, which put a limit 
on the period derivative  of $\dot{P} < 2 \times 10^{-10}$~s~s$^{-1}$. The 
0.5--8.0~keV $Chandra$ ACIS spectrum of the AXP was equally well described by
a blackbody, power-law, or bremsstrahlung continuum model modified by the
interstellar absorption (Muno et al.\ 2006). Analysis with a more complex
model of the magnetar atmosphere suggested that the emission likely
arises in one or more hot spots covering a small fraction of the surface
(Skinner, Perna, and Zhekov, 2006). Search for an infrared counterpart
with the Son of ISAAC instrument on the European Southern Observatory (ESO) 
New Technology Telescope (NTT) yielded negative result within the
\timeform{0.3''} uncertainty in the location of the AXP (Muno et al.\ 2006).

Following the 
detection of an intense ($\sim$10$^{39}$~erg~s$^{-1}$) and short (20~ms) 
burst from CXOU~J164710.2-455216 with the Burst Alert Telescope (BAT) on 
Swift on 2006 September 21 (Krimm et al.\ 2006), the AXP was observed with 
various X-ray observatories. Chandra Target of Opportunity (TOO) observation 
of the AXP on 2006 September 27 showed the presence of 10.61069~s pulsations 
in the ACIS-S event data (Gavriil et al.\ 2006). 
Using the Swift, XMM-Newton, and Chandra data, Israel et al.\ (2007)
obtained a phase-coherent solution for the source pulsations after the
burst. The solution required an exponential component decaying with a
time scale of 1.4~d, which was interpreted to indicate a recovery stage
following a glitch with $\Delta P/P \sim -10^{-4}$. They also
detected a spin-down of $\dot{P} \sim 9 \times 10^{-13}$~s~s$^{-1}$, which
implies a magnetic field strength of $10^{14}$~G\@.
Suzaku performed a TOO observation of the AXP CXOU~J164710.2-455216 on 
2006 September 23--24. The results obtained from the analysis of the Suzaku 
observation are presented in this paper.

\section{Observation}
Following the outburst detected by the BAT, Suzaku performed a TOO 
observation of the AXP CXOU~J164710.2-455216 on 2006 September 23 from
06:52 UT to 04:56 UT next day. The TOO observation was carried out
at ``XIS nominal'' pointing position for effective exposures of 38.8~ks 
with the XIS and 27.7~ks with the HXD\@.  The XIS was operated with ``1/8 
window'' option which gives a time resolution of 1~s, covering a field of 
view of \timeform{17.8'} $\times$ \timeform{2.2'}. 

Suzaku, the fifth Japanese X-ray astronomy satellite, was launched on 2005 
July 10 (Mitsuda et al.\ 2007). It covers 0.2--600~keV energy range with the 
two sets of instruments, X-ray CCDs (X-ray Imaging Spectrometer; XIS) covering 
the soft X-rays in 0.2--12~keV energy range, and the Hard X-ray
Detectors (HXD) which covers 10--70~keV with PIN diodes and 30--600~keV
with GSO scintillators.  
There are 4 XIS (one back illuminated and three front illuminated), each with 
a 1024 $\times$ 1024 pixel X-ray-sensitive CCD detectors at the foci of the 
each of the four X-ray Telescopes (XRT)\@. 
The HXD is a non-imaging instrument that is designed to detect high-energy 
X-rays. It has 16 identical units made up of two types of detectors, silicon 
PIN diodes ($<$70~keV) and GSO crystal scintillator ($>$30~keV). 
For a detailed description of the XIS and HXD detectors, refer to Koyama
et al.\ (2007) and Takahashi et al.\ (2007), respectively.

\section{Analysis and Results}

We used public data (rev-1.2) for the Suzaku TOO observation of the AXP 
CXOU~J164710.2-455216 in present work. For the XIS data reduction, the 
accumulated events were discarded when the telemetry was saturated, data 
rate was low, the satellite was in the South Atlantic Anomaly (SAA), and 
when the source elevation above the earth's limb was below 5$^\circ$ with 
night-earth and below 20$^\circ$ with day-earth. We corrected the known
shift of XIS time assignment in rev-1.2 data (7~s for the current
data).\footnote{http://www.astro.isas.jaxa.jp/suzaku/analysis/xis/timing/}
Applying these conditions, the source spectra were accumulated 
by selecting a circular region of a radius of \timeform{4.3'} 
around the image center; the circle covers 99\% of a point source flux.
Using the same circular region, X-ray light 
curves of 1~s time resolution were also extracted from the XIS event data. 
Because this extraction circle is larger than the optional window, the 
effective extraction region is the intersection of the window and this 
circle. The XIS background spectra were accumulated from the same 
observation by selecting rectangular regions away from the source. 
The response files and effective area files for XIS detectors were 
generated by using the ''xissimarfgen'' and ''xisrmfgen'' task of FTOOLS 
(V6.2). For HXD/PIN data reduction, we used the cleaned event data to 
obtain the source light curve and spectrum. The simulated background 
events were used to estimate the HXD/PIN background (Kokubun et al.\ 2007). 

\begin{figure}
\begin{center}
\rotatebox{-90}{\FigureFile(60mm,60mm){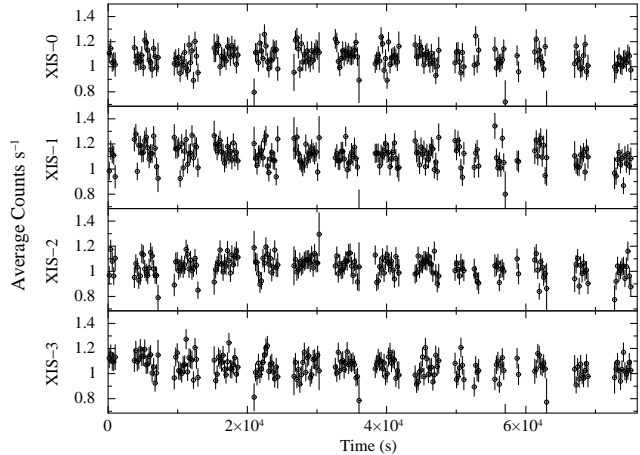}}
\end{center}
\caption{XIS light curves of CXOU~J164710.2--455216 obtained from the
Suzaku TOO observation of the AXP\@. The light curves are plotted for a
binsize of 200 times the spin period of the AXP\@.}
\label{xislc}
\end{figure}

It is found that, during the Suzaku TOO observation, GX~340+0, a bright 
Z-source which is located at about 21 arcmin away from the AXP, contaminated
the HXD data. The RXTE/ASM monitoring data of GX~340+0 shows a more or 
less constant flux around 30 counts s$^{-1}$. Though GX~340+0 was outside 
the field-of-view of the XIS, it was well within the HXD field-of-view. 
Considering a steep power-law spectrum at hard X-rays for AXPs, a roughly 
estimated 10--50 keV flux of CXOU~J164710.2--455216 is found to be a few 
orders of magnitude lower than that of the observed HXD/PIN flux in above 
energy band. This suggests that the HXD data of the Suzaku observation of 
the AXP is significantly contaminated from the contribution of the nearby 
hard X-ray source GX~340+0. In fact, we could not detect significant X-ray 
pulsation at 10.6106 s in the HXD/PIN data. Therefore, in the present work, 
we use only the XIS data.

\subsection{Timing Analysis}
For the timing analysis, a barycentric correction was applied to 
the arrival times of the X-ray photons using the ``aebarycen'' task of
FTOOLS\@. As described above, light curves with 
a time resolution of 1~s were extracted from XIS (0.2--12~keV) event data. 
Light curves with a binsize of 1, 10, 25, 50, 100, and 200 times the pulse 
period (10.61063~s) were investigated without any success in finding
significant variations of the count rate. 
The light curve with 200 times the pulse period 
obtained from all four XIS detectors during the entire observation are shown in
Figure~\ref{xislc}. 

\begin{figure}
\begin{center}
\rotatebox{-90}{\FigureFile(70mm,70mm){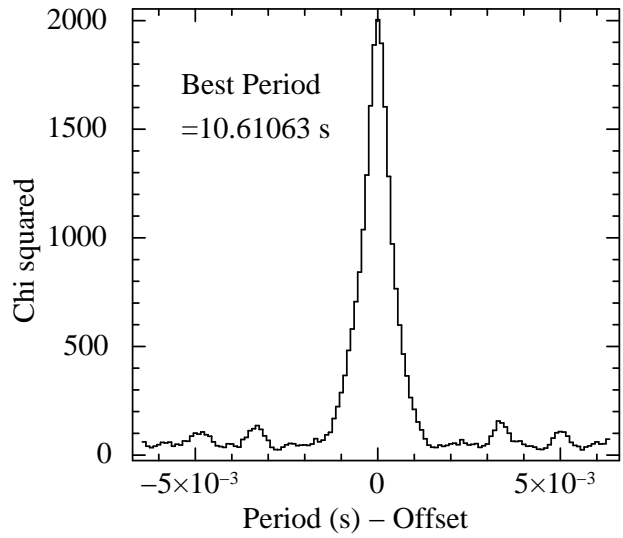}}
\end{center}
\caption{Results of the epoch-folding analysis on the XIS light curve (all four
XIS light curves added together) of CXOU~J164710.2--455216 obtained from the
Suzaku TOO observation of the AXP. }
\label{pp}
\end{figure}

\begin{figure}
\begin{center}
\rotatebox{-90}{\FigureFile(60mm,60mm){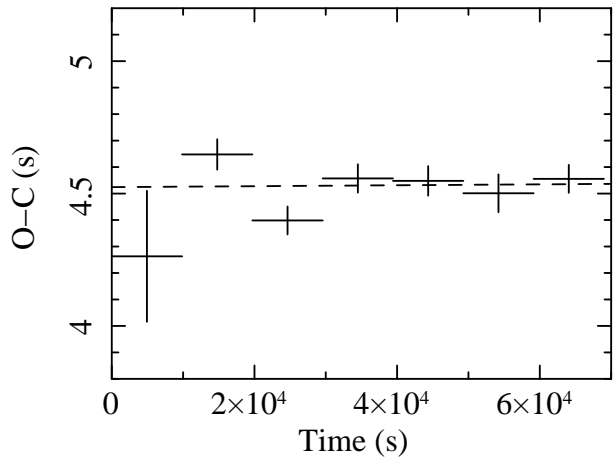}}
\end{center}
\caption{Relative arrival times of the main pulse are plotted when a constant
period of 10.61063~s and an epoch of MJD 54001.000081 are assumed.
See text for details of the calculation.
This plot represents the so-called O--C curve.
Broken line is the best-fit linear function to the data, whose slope
represents offset of the assumed pulse period from the true one.}
\label{occurve}
\end{figure}

\begin{table*}
\begin{center}
\caption{Parameters of the pulse profile}
\begin{tabular}{lccc}
\hline
\hline
Component & 0.2--1.8 keV & 1.8--6.0 keV & 6--12 keV \\
\hline
\hline
Constant         & 0.87 $\pm$ 0.03 & 0.86 $\pm$ 0.02 &  (0.86) \\
Main gaussian & & & \\
\qquad   Centroid      & 0.42 $\pm$ 0.01 & 0.43 $\pm$ 0.01 &  0.40 $\pm$ 0.03 \\
\qquad   Width         & 0.09 $\pm$ 0.02 & 0.09 $\pm$ 0.01 &  0.16 $\pm$ 0.03 \\
\qquad   Normalization & 0.36 $\pm$ 0.05 & 0.41 $\pm$ 0.02 &  0.30 $\pm$ 0.05 \\
2nd gaussian & & & \\
\qquad   Centroid      & 0.76 $\pm$ 0.02 & 0.74 $\pm$ 0.01 &  (0.74) \\
\qquad   Width         & 0.06 $\pm$ 0.02 & 0.06 $\pm$ 0.01 &  (0.06) \\
\qquad   Normalization & 0.15 $\pm$ 0.05 & 0.20 $\pm$ 0.02 &  $<$0.13 \\
3rd gaussian & & & \\
\qquad   Centroid      & 1.02 $\pm$ 0.02 & 1.03 $\pm$ 0.01 &  (1.03) \\
\qquad   Width         & 0.08 $\pm$ 0.03 & 0.07 $\pm$ 0.02 &  (0.07) \\
\qquad   Normalization & 0.14 $\pm$ 0.05 & 0.09 $\pm$ 0.02 &  $<$0.12 \\
\hline
\multicolumn{4}{p{10cm}}{
Note: Gaussian is defined as $N*\exp\{-(x-c)^2/2w^2\}$, where
      $N$ is a normalization, $c$ is a centroid, and $w$ is a width.
      Errors are in 90\% confidence limit.
      Values in the parenthesis are fixed in the fitting.
}
\end{tabular}
\label{pp_par}
\end{center}
\end{table*}

To determine the pulse period of CXOU~J164710.2--455216, all four XIS light
curves were added together to improve the statistics. As the XISs have very
low backgrounds, background subtraction from the light curves was not done.
Pulse folding and $\chi^2$ maximization method was first applied to the
added XIS light curve using the XRONOS task ``efsearch''.
The result is shown in figure~\ref{pp}. 
This analysis yielded the pulse period to be 10.61063~s.
To improve the estimation of the pulse period, we next
applied a phase fitting technique to the XIS data.
We divided the XIS light curve into 8 segments
and calculated a folded pulse profile for
each segment with a common epoch (MJD 54001.000081)
and a period (10.61063~s).
All the segments have a consistent profile with 3 peaks,
but the statistics was poor for the 8th segment, which
was not used for the subsequent analysis.
We determined the phases of the main peak by
fitting a gaussian to the profile.
The phases are converted to the relative arrival times of the
pulse by multiplying the pulse period.
The results are plotted in figure~\ref{occurve} (this is a so-called
O--C curve).
The slope of the plot indicates the adjustment to the
trial period used for the calculation of the folded profile.
The slope was found to be $0.2\pm1.3\times10^{-6}$.
This means that the pulse period of the AXP was 10.61063(2)~s,
where the error corresponds to 90\% confidence limit. This pulse
period is consistent with that obtained from XMM-Newton observation on
2006 September 22 (Muno et al.\ 2007). 

\begin{figure}
\begin{center}
\rotatebox{-90}{\FigureFile(55mm,45mm){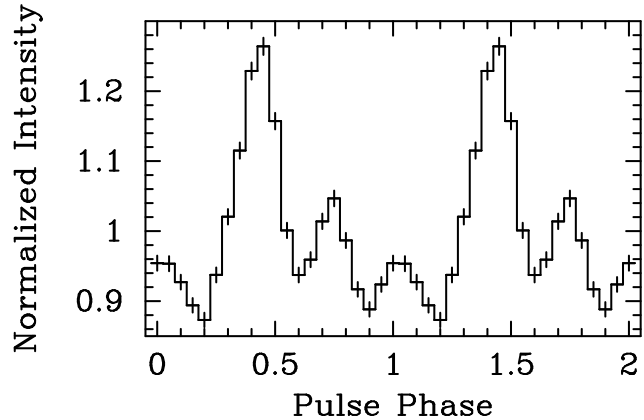}}
\end{center}
\caption{The pulse profile of CXOU~J164710.2--455216 obtained from the
Suzaku TOO observation of the AXP\@. The error bars represent 1$\sigma$ 
uncertainties. Two pulses are shown for clarity.}
\label{xispp}
\end{figure}

The pulse profile obtained from the added XIS light curve of the 
Suzaku observation of the AXP is shown in figure~\ref{xispp}. 
The RMS fractional amplitude of the pulse was found to be $\sim$11\%.
From the figure, it is observed that the shape of the 
pulse profile in the XIS energy band (0.2 -- 12~keV) is not sinusoidal in 
nature, rather it has a three peaked profile. Such a three-peaked
profile was also observed by XMM-Newton on September 22 (Muno et
al.\ 2007), and Chandra on September 27 (Gavriil et al.\ 2006). 
Possibly this is the only AXP which 
shows a three peaked pulse profile. To investigate the energy dependence 
of the pulse profile of the AXP, we generated light curves in different 
energy bands from all four XIS event data. The light curves are folded 
with the pulse period using the XRONOS task ``efold'' and the
corresponding pulse profiles are shown in Figure~\ref{erpp}. 
The RMS fractional amplitude tends to decrease slightly toward the
higher energies: $\sim$11\% in 1.8--6.0~keV while $\sim$8\% 
in 6--12~keV\@.
The energy resolved pulse profiles in XIS energy 
band are found to be different, three peaked profiles with different 
amplitudes in 0.2--6.0~keV energy band and a single peaked profile in 
6--12~keV energy band. 

We carried out a model fitting to the pulse profiles to study the
energy dependence of the profile quantitatively.
We adopted a model consisting of a constant and three gaussian 
functions to represent the pulse profile.
Results of the model fitting was summarized in table~\ref{pp_par}.
We can see in the table that all the model parameters are same
within the errors between 0.2--1.8~keV and 1.8--6.0~keV\@.
On the other hand, the profile in 6--12~keV seems to be
intrinsically different from those in the lower energy bands.
The main peak is significantly broader in 6--12~keV\@.
The 2nd peak, even if present in 6--12~keV, should be
significantly smaller than that below 6~keV\@.
Because of these two differences, the pulse profile in 6--12~keV
looks more or less sinusoidal.
We consider from these differences that the pulse profile in
6--12~keV is intrinsically different from that below 6~keV\@.

\subsection{Spectral Analysis}

\subsubsection{Pulse phase averaged spectroscopy}
Source and background spectra were extracted from the XIS event files 
as described in the beginning of the section. The response files and 
effective area files for XIS detectors were generated by using the 
''xissimarfgen'' and ''xisrmfgen'' task. Events were selected in the 
energy ranges of 0.8--10.0~keV for the back illuminated and front 
illuminated CCDs. After appropriate background subtraction, 
simultaneous spectral fitting was done for the XIS BI and FI spectra
with XSPEC V11. 
All the spectral parameters other than the relative instrument 
normalizations, were tied for the BI and FI detectors. Because an
artificial structure is known to exist in the XIS spectra around the Si
edge, we ignored energy bins between 1.75--1.85~keV in the spectral
analysis.  The energy bins below 0.8~keV were also ignored because of
the lack of photons at soft X-rays. 

\begin{figure}[h]
\begin{center}
\rotatebox{-90}{\FigureFile(95mm,95mm){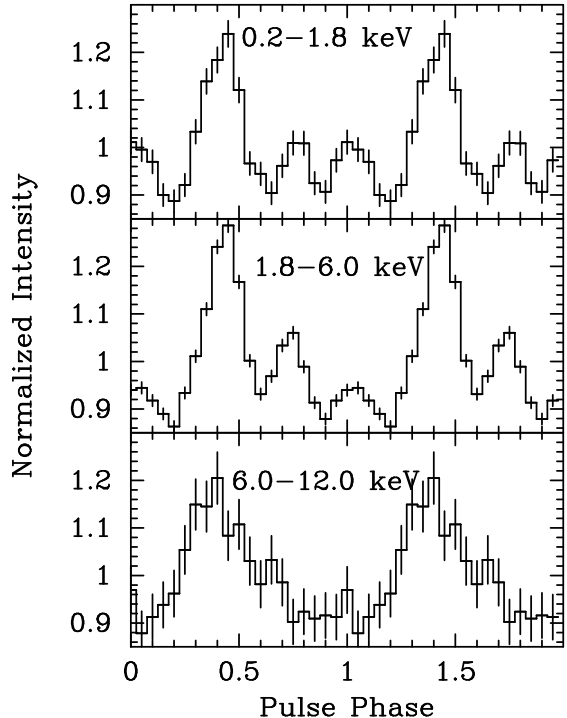}}
\end{center}
\caption{The XIS pulse profiles of CXOU~J164710.2--455216 at different
energy bands showing the presence/absence of the second and third peaks
with energy. No background was subtracted from the folded profiles.
The error bars represent 1 sigma uncertainties.
Two pulses in each panel are shown for clarity.}
\label{erpp}
\end{figure}

\begin{figure}
\begin{center}
\rotatebox{-90}{\FigureFile(70mm,70mm){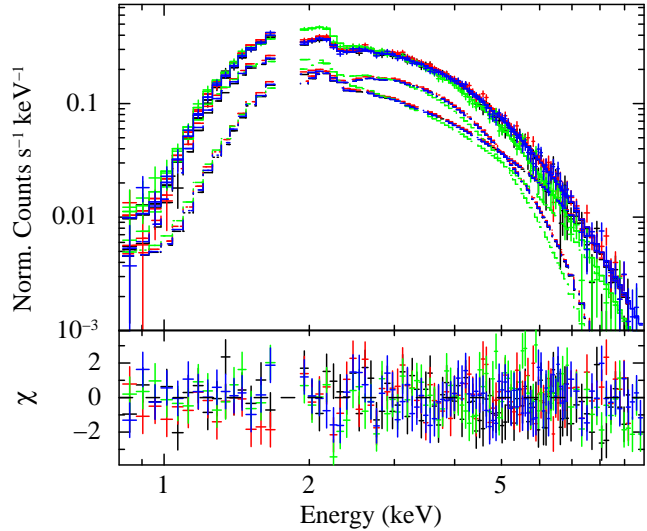}}
\end{center}
\caption{Energy spectrum of CXOU~J164710.2--455216 obtained with the four XIS
detectors of the Suzaku observation, along with the best-fit model comprising
a blackbody component and a power-law continuum model. The bottom panel shows 
the contributions of the residuals to the $\chi^2$ for each energy bin.}
\label{bbpo}
\end{figure}

We tried to fit the XIS spectra of CXOU~J164710.2--455216  using 
a power-law continuum component along with the interstellar absorption. 
Simultaneous spectral fitting of 0.8--10 keV spectrum with above 
model yielded a poor fit with reduced $\chi^2$ of 3.9 for 386 degrees of 
freedom (dof). We have tried to fit the spectra with a blackbody model 
modified with the interstellar absorption. This model improves the spectral 
fitting with a reduced $\chi^2$ of 2.7 for 386 dof. We tried to fit
the spectrum using a model consisting of a blackbody and a power-law component 
with interstellar absorption. This two-component model provided a best-fit to 
the XIS spectra of the AXP with a reduced $\chi^2$ of 1.15 for 384 dof. The spectral 
parameters of the best-fit model obtained from the simultaneous spectral 
fitting are given in table~\ref{spec_par}. The count rate spectra of the 
Suzaku observation is shown in figure~\ref{bbpo} along with the model 
components (top panel) and residuals to the best-fit continuum model 
(bottom panel). Though the estimated absorption column density is found to
be high, the other parameters obtained from the Suzaku observation of 
the AXP are found to agree to that reported from Chandra and XMM-Newton
observations (Gavriil et al.\ 2006; Muno et al.\ 2007). 

\begin{figure}
\begin{center}
\rotatebox{-90}{\FigureFile(70mm,70mm){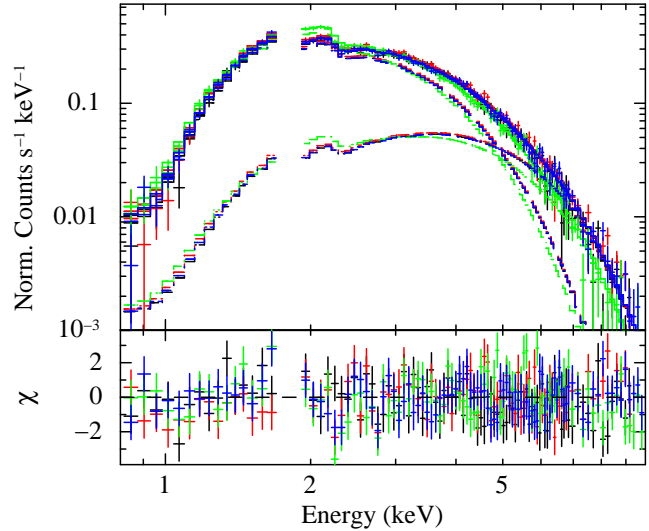}}
\end{center}
\caption{Energy spectrum of CXOU~J164710.2--455216 obtained with the four XISs
in the Suzaku observation, along with the best-fit model comprising
two blackbody components as continuum model. The bottom panel shows
the contributions of the residuals to the $\chi^2$ for each energy bin.}
\label{2bb}
\end{figure}

Though the two component model (power-law and blackbody components) fits very
well to the 0.8--10.0~keV XIS spectra, this model is not compatible with the
energy dependence of the pulse profile. From figure~\ref{erpp},
the pulse profile is three peaked at soft X-rays, but tends to be 
single peaked at hard X-rays. 
On the other hand, it is found that the power-law component dominates 
below 2 keV and above 5 keV and the blackbody component dominates 
in 2--5~keV energy band (Figure~\ref{bbpo}). 
This means that the pulse profile should be
similar below 2~keV and above 5~keV, which is not the case. 
Therefore, the model with a blackbody and a power-law as model 
components is not favored.
Following this, we tried to fit the 1--10~keV spectra with a model 
consisting of two blackbody components. This model fits the data very well 
with a reduced $\chi^2$ of 1.19 for 384 dof. The best-fit parameters are 
given in table~\ref{spec_par} and the spectra with the fitted model components
are shown in figure~\ref{2bb}. The two blackbody model was reported to
fit the XMM-Newton data (Muno et al.\ 2007) and the Swift data (Israel et
al.\ 2007) just after the burst. The best-fit parameters obtained with
Suzaku are comparable to those obtained by these satellites.

Addition of a narrow Gaussian function at 6.4~keV (for iron K$_\alpha$ 
fluorescence line) did not yield any change in the value of the reduced 
$\chi^2$ and the spectral parameters. The corresponding equivalent width 
is found to be about 6~eV and the flux of the emission line is estimated 
to be 8.4$\times$10$^{-15}$  ergs cm$^{-2}$ s$^{-1}$ which is about three 
orders of magnitude lower than the source flux. This suggests the lack of
reprocessing matter surrounding the neutron star to produce iron 
fluorescence emission.

\begin{table*}
\begin{center}
\caption{Spectral parameters for CXOU~J164710.2--455216}
\begin{tabular}{lcc}
\hline
\hline
Parameter	&Blackbody+Power-law       & Blackbody+Blackbody\\
\hline
\hline
N$_H$ (10$^{22}$ atoms cm$^{-2}$)    &2.55$\pm$0.01     &1.73$\pm$0.03\\
$kT_{BB1}$ (keV)                     &0.67$\pm$0.01     &0.61$\pm$0.01\\
$kT_{BB2}$ (keV)                     &------            &1.22$\pm$0.06\\
Power-law index ($\Gamma_1$)         &3.14$\pm$0.08       &-----\\
Blackbody flux (F$_{BB1}$)$^a$       &1.3$\pm$0.1       &1.8$\pm$0.1\\
Power-law flux (F$_{PO}$)$^a$	     &1.3$\pm$0.1       &-----\\
Blackbody flux(F$_{BB2}$)$^a$        &------            &0.8$\pm$0.1\\
Total source flux$^a$                &2.6$\pm$0.1       &2.6$\pm$0.1\\
Reduced $\chi^2$		     &1.15 (384 dof)   &1.19 (384 dof)\\
\hline
\multicolumn{3}{l}{
Note: Errors are defined in 1$\sigma$ confidence limit.}\\
\multicolumn{3}{p{12cm}}{
$^a$ : Flux (in 10$^{-11}$ ergs cm$^{-2}$ s$^{-1}$) is estimated in 
1--10~keV energy range without the correction of the absorption.
}
\end{tabular}
\label{spec_par}
\end{center}
\end{table*}

\subsubsection{Pulse phase resolved spectroscopy}

Different type of pulse profiles of CXOU~J164710.2--455216 in different 
energy bands prompted us to make phase resolved spectral analysis of 
the Suzaku observation of the AXP\@. To investigate the changes in 
the spectral parameters at soft X-rays at different pulse phases, 
the source spectra were accumulated into 10 pulse phase bins by
applying phase filtering in the FTOOLS task XSELECT\@. The XIS 
background spectra and response matrices used for the phase averaged 
spectroscopy, were also used for the phase resolved spectroscopy.
Simultaneous spectral fitting was done in the 1--10~keV energy band.

\begin{figure*}
\begin{center}
\rotatebox{-90}{\FigureFile(120mm,120mm){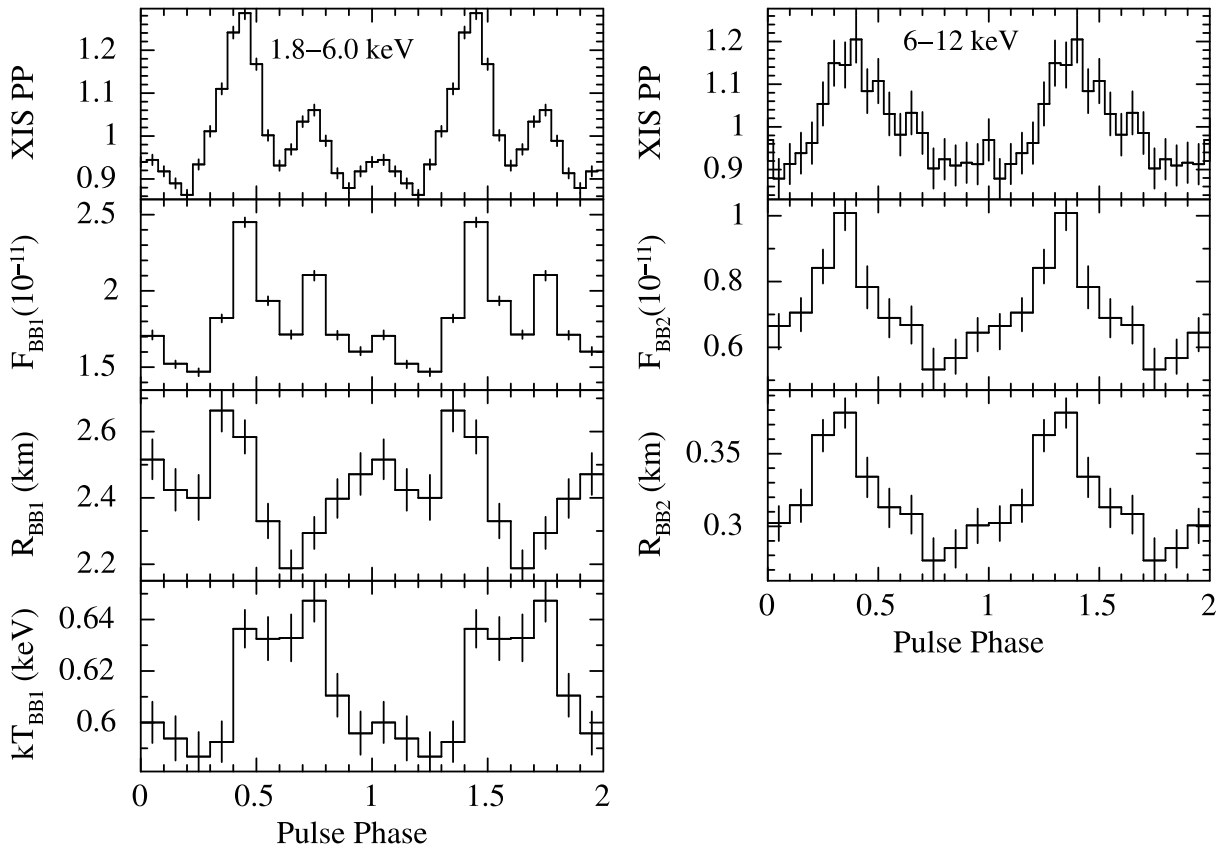}}
\end{center}
\caption{Spectral parameters obtained from the pulse phase resolved
spectroscopy of Suzaku observation of CXOU~J164710.2--455216. In the 
figure, the change in the soft blackbody flux (Flux$_{BB1}$), and 
second blackbody flux (Flux$_{BB2}$) over pulse phases are shown along 
with the blackbody temperature kT$_{BB1}$ and the radii (assuming the 
source distance to be 5~kpc) of the blackbody emitting regions (R$_{BB1}$ 
and R$_{BB2}$) with 1$\sigma$ errors. The blackbody flux is in the units 
of 10$^{-11}$~ergs~cm$^{-2}$~s$^{-1}$. The XIS pulse profiles (XIS PP) in
1.8--6.0~keV and 6--12~keV energy bands are also shown in the left and right
top panels respectively.} 
\label{phrs}
\end{figure*}

\begin{figure*}
\begin{center}
\rotatebox{0}{\FigureFile(90mm,90mm){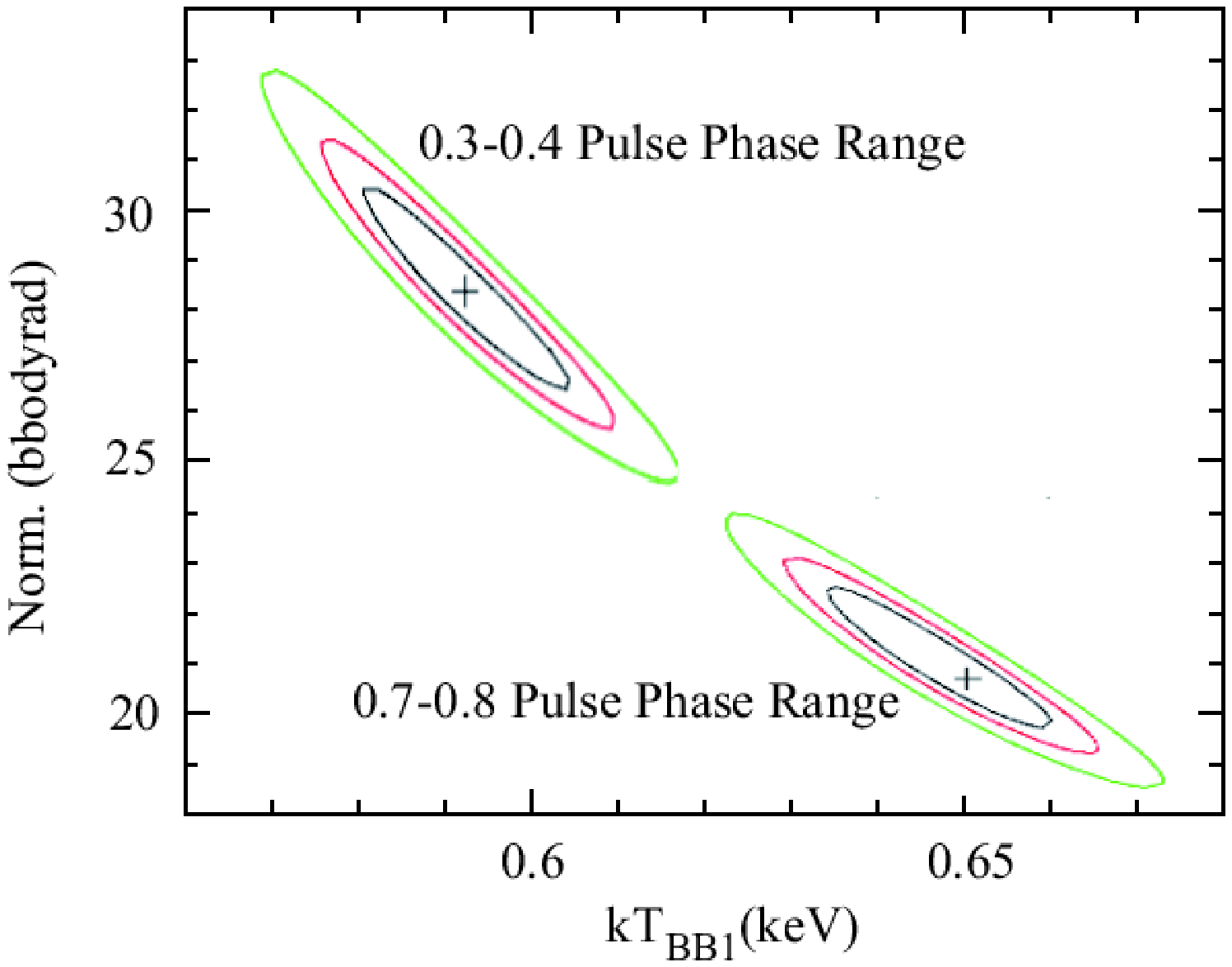}}
\end{center}
\caption{Confidence contours between the normalization and the
 temperature of the soft blackbody component ($kT_{BB1}$) obtained
 for the energy spectra in 0.3--0.4 and 0.7--0.8 pulse phase ranges.
Normalization is defined as $R_{BB1}^2/D_{10}^2$, where $D_{10}$ is the
 source distance in unit of 10 kpc.
Three contours correspond to 68\%, 90\%, and 99\% confidence limits 
respectively.}
\label{cont}
\end{figure*}

The phase resolved spectra were fitted with the same two blackbody 
component model used to describe the phase averaged spectrum. 
The value of the absorption column density (N$_H$) and the temperature
of the second blackbody component (kT$_2$) did not show any significant 
variability over the pulse phase. We kept fixed the values of N$_H$ 
and kT$_2$ to that of the phase average values during the spectral 
fitting of the phase resolved spectra. The parameters obtained from the 
spectral fitting to the XIS phase resolved spectra are shown in 
figure~\ref{phrs} along with the XIS pulse profiles at the top panels,
and are listed in table~\ref{spec_phase_par}.
As the soft blackbody component dominates the spectrum at soft 
X-rays and the second blackbody dominates the spectrum at $>$5~keV,
the change in the estimated flux of the two blackbody components 
over the pulse period is expected to resemble the pulse profiles at 
corresponding energy bands. From the figure, it is found that the soft 
blackbody (marked as BB1 in figure) flux profile shows two prominent 
and narrow peaks and a third small peak at same phases as in the
0.2--12~keV pulse profile of the AXP\@. On the otherhand, the second
blackbody (BB2) flux shows a single peak profile which is in phase with
the 6--12~keV pulse profile of the pulsar. 
The flux of the soft blackbody takes a maximum in phase 0.4--0.5,
whereas that of the second blackbody takes a maximum in phase 0.3--0.4.
The blackbody temperature (kT$_{BB1}$) rises from minimum 
($\sim$0.59~keV) at phase 0.2--0.3 to maximum ($\sim$0.65~keV) at
0.7--0.8 phase.  
The radius of the soft blackbody component varies between 2.2~km and
2.7~km and peaks in phase 0.3--0.4. However, the blackbody radius of the
second component varies between 0.28~km and 0.38~km, again peaking in
phase 0.3--0.4.  
The radius of the blackbody emitting region is estimated assuming the 
source distance as 5~kpc. To check the acceptable ranges of the blackbody 
temperature ($kT_{BB1}$) and the radius of the blackbody emitting region, 
confidence contours were plotted between $kT_{BB1}$ and blackbody 
normalization for 0.3--0.4 and 0.7--0.8 phase ranges
(figure~\ref{cont}).
For the phase 0.3--0.4, the acceptable range
of $kT_{BB1}$ and the blackbody normalization (N$_{BB1}$) are found to 
be 0.57--0.62~keV and 24--33, respectively. 
However, for phase 0.3--0.4, the acceptable ranges of $kT_{BB1}$ and
N$_{BB1}$ are found to be 0.62--0.67~keV and 18.5--24.0, respectively, for
99\% confidence limit.  The confidence  
contours for above phase ranges show that the change in the soft blackbody 
temperature and the radius of the soft blackbody emitting region over pulse
phase of the AXP is genuine.

\begin{table*}
\begin{center}
\caption{Spectral parameters in each pulse phase}
\begin{tabular}{llllll}
\hline
\hline
Phase & F$_{BB1}^*$ & F$_{BB2}^*$ & R$_{BB1}^\dagger$ & R$_{BB2}^\dagger$ & kT$_{BB1}^\ddag$\\
\hline
\hline
0.0 -- 0.1 &    $1.71_{-0.02}^{+0.04}$ &        $0.67_{-0.07}^{+0.04}$ &        $2.52\pm0.06$ &$0.30\pm0.01$ & $0.60\pm0.01$ \\
0.1 -- 0.2 &    $1.52\pm0.03$ & $0.71_{-0.06}^{+0.04}$ &        $2.42\pm0.06$ & $0.31\pm0.01$ &$0.59\pm0.01$ \\
0.2 -- 0.3 &    $1.47\pm0.03$ & $0.84\pm0.06$ & $2.40\pm0.07$ & $0.36\pm0.01$ & $0.59\pm0.01$ \\
0.3 -- 0.4 &    $1.82\pm0.03$ & $1.01\pm0.06$ & $2.66\pm0.06$ & $0.38\pm0.01$ & $0.59\pm0.01$ \\
0.4 -- 0.5 &    $2.45\pm0.03$ & $0.78\pm0.07$ & $2.58\pm0.05$ & $0.33\pm0.01$ & $0.64\pm0.01$ \\
0.5 -- 0.6 &    $1.94\pm0.03$ & $0.69\pm0.06$ & $2.33\pm0.05$ & $0.31\pm0.01$ & $0.63\pm0.01$ \\
0.6 -- 0.7 &    $1.71\pm0.03$ & $0.67\pm0.06$ & $2.19\pm0.05$ & $0.31\pm0.01$ & $0.63\pm0.01$ \\
0.7 -- 0.8 &    $2.10\pm0.04$ & $0.53\pm0.06$ & $2.29\pm0.05$ & $0.28\pm0.02$ & $0.65\pm0.01$ \\
0.8 -- 0.9 &    $1.71\pm0.03$ & $0.57\pm0.06$ & $2.40\pm0.06$ & $0.29\pm0.01$ & $0.61\pm0.01$ \\
0.9 -- 1.0 &    $1.60\pm0.03$ & $0.64\pm0.06$ & $2.47\pm0.06$ & $0.30\pm0.01$ & $0.60\pm0.01$ \\
\hline
\multicolumn{6}{l}{
Note: Errors are defined in 1$\sigma$ confidence limit.}\\
\multicolumn{6}{p{12cm}}{
$^*$Flux (in 10$^{-11}$~ergs~cm$^{-2}$~s$^{-1}$) is estimated in 
1--10~keV energy range and is not corrected for absorption.}\\
\multicolumn{6}{p{12cm}}{
$^\dagger$In unit of km for the assumed source distance of 5 kpc.}\\
\multicolumn{6}{p{12cm}}{
$^\ddag$In unit of keV.}
\end{tabular}
\label{spec_phase_par}
\end{center}
\end{table*}

\section{Discussion}

The observed changes in the AXP CXOU~J164710.2--455216 before and 
after the intense and short burst in 2006 September is remarkable. 
The XMM-Newton observation on 2006 September 22 
and the Suzaku observation on 2006 September 23 summaries 
various changes in the AXP, when compared with the pre-burst
data of XMM-Newton on September 16 (Muno et al.\ 2007).
The pulse profile of the AXP was a single peaked before the burst, 
which changed to a three peaked profile on September 22. 
These peaks were seen in 0.5--7.0~keV range. 
Although only the pulse profile below 7~keV was presented
in Muno et al.\ (2007), current Suzaku data show that the pulse
profile is single peaked at higher energy band, in 6--12 keV\@.
Because the spectral parameters are comparable between the
XMM-Newton data and the Suzaku data, we consider that the
nature of the pulse profile, especially the energy dependence, 
is basically same between these two sets of observations.
Though the pulse profiles of some of the AXPs
show changes before and after the burst (Kaspi et al.\ 2003), 
the change from a single peak profile to a three peak profile 
is unique to CXOU~J164710.2--455216. 

Before the intense and short burst, the spectrum was described by a single 
blackbody component and the pulse profile was single peaked (Muno et al.\ 
2007). The hardening of the source spectrum (after the burst) was seen from 
the XMM-Newton observations which was explained by adding a power-law 
component with an index of $\sim$2.0 to the blackbody component to describe 
the spectrum. The pulse profile also changed to a three peaked profile as 
seen from XMM-Newton and Suzaku observations. However, energy resolved 
pulse profiles of Suzaku observation shows that the hard X-ray (6--12~keV)
pulse profile is similar to that of the pre-burst profile obtained from 
XMM-Newton whereas the soft X-ray pulse profile is found to be three
peaked. This energy dependent pulse profile of Suzaku observation ruled 
out the model with blackbody and power-law as spectral components and favored
a model with two blackbody components. The phase resolved spectroscopy of 
Suzaku observation also showed that the change in the soft blackbody flux 
over pulse phase shows two narrow and intense peaks and a third minor peak 
exactly at the same phases of the peaks in the pulse profile. The second 
blackbody flux, however, shows only a single peak which is similar to the 
6--12~keV pulse profile. These findings suggest that the single blackbody 
component which described both the spectrum and the pulse profile before 
the outburst is also required to describe the source properties after the 
outburst with an increased temperature and hence at hard X-ray band. The 
three peaked profile, seen after the outburst, is interpreted due to 
the emergence of a new spectral component which dominates at soft X-rays. 
This new spectral component may be originated from the giant flares in the 
AXPs/SGRs during which a significant amount of the magnetic energy is 
released.

The multi-peaked pulse profiles in AXPs/SGRs are attributed to the presence 
of multi-pole structures of the external magnetic fields. The rearrangement 
of these multi-pole magnetic fields changes the shape of the observed pulse 
profiles in these objects. Muno et al.\ (2007) suggested that the observed 
changes in pulse profiles and spectral properties of the AXP before and after 
the burst are due to a change in the distribution of currents in the 
magnetosphere. This change in magnetosphere is triggered by a plastic 
motions in the crust of the neutron star. The observed three-peaked profile
in the AXP CXOU~J164710.2--455216 suggests that three magnetic field foot
points developed in response to the change in magnetoshepre. Above the foot 
points, there might be hot plasma columns, like an accretion column of the 
accretion powered pulsars. The temperatures distribution is hotter temperature 
at the lower altitude and lower temperature at the higher altitude. The 
electron-positron plasma within the column cannot move in the direction 
perpendicular to the magnetic filed. But those can move in the direction 
parallel to the magnetic field. Therefore, the X-ray photons pass through 
along with the magnetic field can escape freely. However, the X-ray photon 
passing in transversal direction to the magnetic field are restricted by 
the electron positron scattering. In that case, black body radiation from 
such foot point is beamed into the magnetic filed direction. This can explain 
the narrow three-peaked profile of the blackbody flux in
CXOU~J164710.2--455216. 


\section*{Acknowledgments}
The authors would like to thank all the members of the Suzaku Science 
Working Group for their contributions in the instrument preparation, 
spacecraft operation, software development, and in-orbit instrumental 
calibration.   SN acknowledges the support by JSPS (Japan Society for
the Promotion of Science) post doctoral fellowship for foreign
researchers (P05249).  This work was partially supported by grant-in-aid
for JSPS fellows (1705249).

\end{document}